\documentstyle[11pt,twoside,colloqOHP2010,epsfig]{article}
\begin{document}

\title{Improving Transit Predictions of Known Exoplanets with TERMS}

\author{S.R. Kane\altaffilmark{1}, D. Ciardi\altaffilmark{1},
  D. Fischer\altaffilmark{2}, G. Henry\altaffilmark{3},
  A. Howard\altaffilmark{4}, E. Jensen\altaffilmark{5},
  G. Laughlin\altaffilmark{6}, S. Mahadevan\altaffilmark{7}, K. von
  Braun\altaffilmark{1}, J. Wright\altaffilmark{7}}
\altaffiltext{1}{NASA Exoplanet Science Institute, Caltech,
  Pasadena, CA, 91125, USA}
\altaffiltext{2}{Dept of Astronomy, Yale University, New Haven, CT,
  06520, USA}
\altaffiltext{3}{Tennessee State University, Nashville, TN, 37209, USA}
\altaffiltext{4}{Dept of Astronomy, University of California,
  Berkeley, CA, 94720, USA}
\altaffiltext{5}{Dept of Physics \& Astronomy,
  Swarthmore College, Swarthmore, PA, 19081, USA}
\altaffiltext{6}{UCO/Lick Observatory, University of California, Santa
  Cruz, CA, 95064, USA}
\altaffiltext{7}{Dept of Astronomy \& Astrophysics, Pennsylvania State
  University, University Park, PA, 16802, USA}

\begin{abstract}

Transiting planet discoveries have largely been restricted to the
short-period or low-periastron distance regimes due to the bias
inherent in the geometric transit probability. Through the refinement
of planetary orbital parameters, and hence reducing the size of
transit windows, long-period planets become feasible targets for
photometric follow-up. Here we describe the TERMS project that is
monitoring these host stars at predicted transit times.

\end{abstract}


\section{Introduction}

Monitoring known radial velocity (RV) planets at predicted transit
times, particularly those planets in relatively eccentric orbits,
presents an avenue through which to investigate the mass-radius
relationship of exoplanets into unexplored regions of
period/periastron space beyond (Kane \& von Braun 2008, 2009). Here we
describe techniques for refining ephemerides and performing follow-up
observations (Kane et al. 2009). These methods are used by the Transit
Ephemeris Refinement and Monitoring Survey (TERMS).

\section{Transit Ephemerides}

The transit window as described here is defined as a specific time
period during which a complete transit (including ingress and egress)
could occur for a specified planet. The size of a transit window will
increase with time due to the uncertainties in the fit parameters,
and thus follow-up of the transit window as soon as possible after
discovery is optimal. Figure 1 (left panel) shows the size of the
transit window for a sample of 245 exoplanets. The transit windows of
the short-period planets tend to be significantly smaller than those
of long-period planets since, at the time of discovery, many more
orbits have been monitored to provide a robust estimate of the orbital
period. TERMS chooses targets that have small transit windows,
medium-long periods, and a relatively high probability of transiting
the host star.

\begin{figure*}
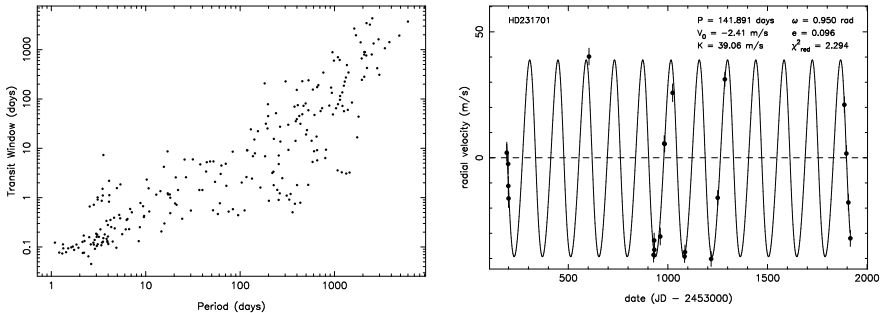

  \begin{center}
    \begin{tabular}{cc}
      \epsfig{angle=270,width=5.6cm,file=f01a.eps} &
      \epsfig{angle=270,width=5.6cm,file=f01b.eps}
    \end{tabular}
  \end{center}
  \caption{Left: Calculated transit windows for a subset of the known
    exoplanets. Right: Published RV data for HD~231701b with four
    additional simulated measurements.}
\end{figure*}


\section{Refining the Ephemerides}

The transit ephemeris for a particular planet can often be
significantly improved with the addition of a handful of
high-precision RV data. For example, the planet orbiting the star
HD~231701 (Fischer et al. 2007) has a current transit window of
$\sim$82 days based upon the discovery data.  The addition of four
subsequent measurements as shown in Figure 1 (right panel) would
improve both the precision of the period and time of periastron
passage, resulting in a reduction of the transit window to 3.7 days -
a factor of almost 25! Through selective observations at optimal
times, we produce viable targets for photometric follow-up.


\section{TERMS Results}

A considerable number of high transit probability targets are
difficult to monitor adequately during their transit windows because
the uncertainties in the predicted transit mid-points are too
high. The acquisition of a handful of new RV measurements at carefully
optimised times can reduce the size of a transit window by an order of
magnitude. This is described in more detail by Kane et al. (2009).

\begin{figure}[ht]
  \begin{center}
    \epsfig{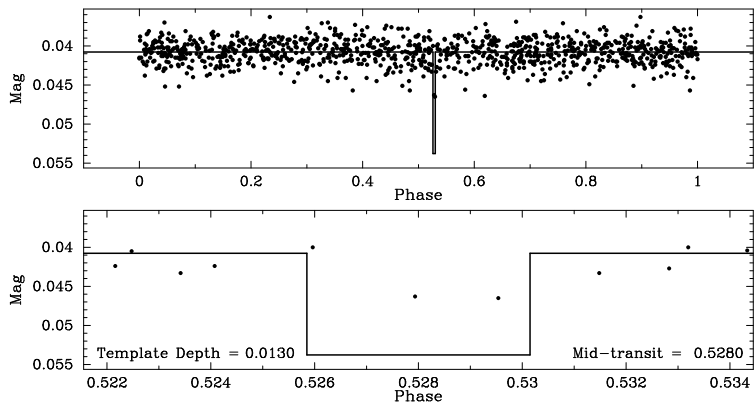}
    \caption{Two panels showing the ruling out of a transit by the
      planet orbiting HD~114762 using data from the T10 APT.}
  \end{center}
\end{figure}

Figure 2 presents data acquired with the T10 0.8m Automated
Photoelectric Telescope (APT) during a predicted transit time of
HD~114762b. The transit window was refined to less than a day using
Lick RV data, and the transit of this planet was subsequently ruled
out. The observations from this survey will lead to improved exoplanet
orbital parameters and ephemerides even without an eventual transit
detection. The results from TERMS will provide a complementary dataset
to the fainter magnitude range of the {\it Kepler} mission, expected
to discover many intermediate to long-period transiting planets.


\end{document}